\documentclass[aps,prd,nopacs,floatfix,notitlepage,nofootinbib,twocolumn,a4paper,longbibliography]{revtex4-1}

\usepackage{amsfonts,amsmath,wasysym,epsfig,graphicx,verbatim,color,subfigure,graphicx,bm,mathrsfs,lipsum,hyperref,cleveref}
\usepackage{booktabs}
\usepackage[normalem]{ulem}  
\usepackage{multirow}
\usepackage{xcolor}

\usepackage[T1]{fontenc}
\usepackage[utf8]{inputenc}
\usepackage[english]{babel}

\usepackage{dcolumn}
\usepackage{tabularx}
\usepackage{makecell}
\usepackage{siunitx}
\usepackage{svg}
\usepackage{hyperref}

\setcellgapes{3pt}

\begin{document}

\newcommand{\OKC}{The Oskar Klein Centre, Department of Astronomy, Stockholm University, AlbaNova, SE-10691 Stockholm,
Sweden}

\newcommand{\HU}{{Hamburger Sternwarte, Gojenbergsweg 112, D-21029 Hamburg, Germany}}

\newcommand{\SNU}{{Department of Physics, School of Natural Sciences, Shiv Nadar Institution of Eminence, NH-91, Tehsil Dadri, Gautam Buddha Nagar, Uttar Pradesh-201314, India}}

\title{Fast and Accurate Prediction of Neutron Star Structure with Deep Neural Networks}

\author{Kaushikk V N$^{\rm 1}$, Bhaskar Biswas$^{\rm 1, \rm 2}$, Stephan Rosswog$^{\rm 1, \rm 3}$}
\affiliation{$^{\rm 1}$\HU, $^{\rm 2}$\SNU, $^{\rm 3}$\OKC}


\begin{abstract}
Solving the Tolman--Oppenheimer--Volkoff (TOV) equations, together with the tidal 
perturbation equations, for large numbers of equation-of-state (EOS) samples is a major 
computational bottleneck in Bayesian inference of the dense-matter EOS, and this will become increasingly limiting as next-generation observatories deliver far larger and more precise datasets. We develop neural-network surrogates for the forward TOV mapping that predict neutron star mass, radius, and tidal deformability simultaneously and directly from the EOS parameters and central density. We train and compare two architectures: a conventional feedforward network and a residual network, the latter of which, to our knowledge, has not previously been explored for TOV surrogate modeling. Trained on a piecewise polytropic EOS parameter space, both networks reproduce the numerical solutions to high accuracy, with the coefficient of determination exceeding 0.999 for all three observables, while accelerating the evaluation of stellar observables by roughly two orders of magnitude relative to direct numerical integration. We find that both architectures achieve excellent predictive accuracy at the network sizes considered here, with the residual network providing a modest improvement in accuracy over the feedforward network at the expense of slightly longer inference times. The overall performance differences remain small, indicating that a feedforward network already has sufficient capacity for this mapping while residual connections offer only incremental gains. Nevertheless, the residual architecture provides a robust baseline for future extensions to richer EOS parameterizations or higher-dimensional regression tasks. The resulting surrogates are  well-suited to large-scale Bayesian EOS inference and population studies, where repeated TOV evaluations would otherwise dominate the computational cost.
\end{abstract}


\maketitle

\section{INTRODUCTION}\label{sec:Intro}

The nuclear equation of state (EOS), which relates pressure and energy density, remains one of the central open problems in relativistic and nuclear astrophysics ~\cite{Lattimer_2016,Oertel_2017,Baym_2018}. Its behavior is experimentally constrained primarily near the nuclear saturation density, $\rho_0 \equiv 2.7 \times 10^{14}\ \mathrm{g\ cm^{-3}}$ (equivalently, $n_0 \equiv 0.16\ \mathrm{fm^{-3}}$). Measurements of neutron-skin thickness in atomic nuclei constrain the symmetry energy at $\rho_0$~\cite{PREX:2021umo,CREX:2022kgg}, while heavy-ion collisions probe isospin-symmetric matter above $\rho_0$~\cite{Danielewicz:2002pu, Li:2008gp, Sorensen:2023zkk}. Nuclear theory extends these constraints to roughly $2\,n_0$~\cite{Epelbaum:2008ga,Machleidt:2011zz,Hammer:2012id,Hebeler:2020ocj,Drischler:2021kxf}. Matter at the most extreme densities, $6$--$8\,n_0$, but essentially zero temperature, is realized only inside neutron stars (NSs), making them the sole natural laboratory for the high-density EOS~\cite{Lattimer:2004sa,Baym_2018}. The remnant oscillations in the aftermath of a neutron star merger, will additionally allow to probe the properties of {\em hot} nuclear matter \cite{Baiotti:2019sew,ET:2025xjr,biswas2026binaryneutronstarmergers}.

The macroscopic structure of a NS, its mass and radius, is determined by the Tolman--Oppenheimer--Volkoff (TOV) equations~\cite{oppenheimer39,tolman39} which require the EOS as input. The tidal deformability, which characterizes the response of a NS to an external tidal field, is obtained by simultaneously solving the TOV equations alongside the differential equations governing perturbations of the metric~\cite{Hinderer:2007mb, Flanagan:2007ix, Damour:2009vw}. Inverting this relationship, i.e., inferring the EOS from observed NS properties, is the central inference problem of NS astrophysics~\cite{1992ApJ...398..569L}. Several observational channels now contribute to this effort. Precise radio timing of binary pulsars yields NS masses~\cite{Demorest:2010bx,Cromartie:2019kug}. Joint mass--radius measurements are obtained from pulse-profile modeling of X-ray emissions from millisecond pulsars, most notably through the NICER mission~\cite{2016SPIE.9905E..1HG, Bogdanov:2019ixe, Riley:2019yda, Miller:2019cac, Riley:2021pdl, Miller:2021qha}. Gravitational wave (GW) observations of binary neutron star (BNS) mergers provide tidal deformability constraints; GW170817, the first observed BNS merger~\cite{LIGOScientific:2017ync,Abbott:2018wiz}, placed the first GW-derived radius estimates~\cite{Abbott:2018exr}. Together, these multimessenger data are routinely combined in Bayesian inference frameworks to constrain the EOS~\cite{Raaijmakers:2019dks,Traversi:2020aaa,Xie:2019sqb,Biswas:2020puz,Al-Mamun:2020vzu,Dietrich:2020efo,Landry_2020PhRvD.101l3007L,Biswas:2021yge,Miller:2021qha,Biswas:2024hja,Biswas:2025ivu,Biswas:2025tjt}.

Direct EOS parameter inference, however, requires solving the TOV equations coupled with the tidal perturbation equations for every proposed sample, typically $\mathcal{O}(10^5$--$10^6)$ times per analysis. This computational cost is already a bottleneck in current pipelines and will become prohibitive with next-generation instruments. 
Upcoming X-ray telescopes such as eXTP~\cite{eXTP:2018anb} and STROBE-X~\cite{STROBE-XScienceWorkingGroup:2019cyd} and next-generation GW observatories (the Einstein Telescope~\cite{Maggiore:2019uih} and Cosmic Explorer~\cite{Reitze:2019iox}) are projected to detect $\mathcal{O}(10^5)$ BNS mergers annually and measure NS radii to sub-100-meter precision~\cite{Chatziioannou:2021tdi, Walker:2024loo}. Meeting this demand requires inference methods that are both accurate and computationally efficient.

A widely explored strategy is to replace the TOV solver with a machine-learning surrogate. 
Prior work has employed regression methods~\cite{Richter:2023zec, Imam:2023ngm}, feedforward neural networks~\cite{Ferreira:2019bny, Thete:2022drz, Tiwari:2024jui, Reed:2024urq, Magnall:2024ffd, DiClemente:2025pbl, liodis_neural-network-based_2024}, support vector machines~\cite{Ferreira:2019bny}, Gaussian processes and reduced basis methods~\cite{Reed:2024urq}, and dynamic mode decomposition~\cite{Lalit:2024vmu} to emulate the forward TOV mapping, including in modified gravity scenarios where the structure equations themselves are altered~\cite{liodis_neural-network-based_2024}. A complementary class of methods bypasses traditional sampling entirely, using neural networks~\cite{Fujimoto:2017cdo, Fujimoto:2019hxv, Morawski:2020izm, Traversi:2020dho, Krastev:2021reh, Fujimoto:2021zas, Soma:2022qnv, Soma:2022vbb, Farrell:2022lfd, Han:2022sxt, Ferreira:2022nwh, Carvalho:2023ele, Krastev:2023fnh, Wu:2023npy, Chatterjee:2023ecc, Ventagli:2024xsh}, normalizing flows~\cite{Morawski:2022aud, Brandes:2024vhw, Hu:2024lrj}, and transformers~\cite{Goncalves:2022smd} to recover the EOS directly from NS observables. Alternatively, rather than parameterizing the EOS directly, one can parameterize the macroscopic $M$-$R$ and $M$-$\Lambda$ relations and infer NS properties from multimessenger data within a Bayesian framework~\cite{Biswas:2021paf}. Despite their speed advantages, surrogate-based methods share a structural limitation: models must be retrained whenever the EOS parameterization or prior assumptions change, and training can require hours to weeks depending on model complexity~\cite{McGinn:2024nkd, Hu:2024lrj}. An alternative that avoids this bottleneck entirely is to leverage differentiable programming and GPU acceleration to solve the TOV equations on the fly during inference~\cite{Wouters:2025zju}, eliminating the need for pre-trained emulators at the cost of restricting flexibility to gradient-compatible samplers.

In this work, we develop and benchmark neural-network surrogates for the forward TOV 
mapping, including the tidal deformability. We consider two architectures: a conventional feedforward network (FFN), which has been used successfully in previous TOV emulation studies~\cite{Ferreira:2019bny, Thete:2022drz, Tiwari:2024jui, Reed:2024urq}, and a residual network (ResNet), which to our knowledge has not previously been applied to TOV surrogate modeling. Residual architectures have achieved considerable success in other domains by enabling the stable training of deeper networks through shortcut connections~\cite{he2015deepresiduallearningimage}, and the strongly nonlinear dependence of neutron star observables on the EOS makes them a natural candidate to investigate for this task. Using a large dataset of neutron star models generated from a piecewise polytropic EOS parameterization~\cite{read_constraints_2009}, we train both surrogates to predict mass, radius, and tidal deformability simultaneously, and assess their accuracy, training behavior, and computational efficiency against direct numerical integration of the TOV equations.
The resulting surrogate models provide orders-of-magnitude reductions in evaluation time, making them attractive for computationally intensive applications such as Bayesian inference, parameter estimation, and population studies that require repeated TOV evaluations.

The remainder of this paper is organized as follows. Section \ref{sec:TOV} summarizes  the Tolman–Oppenheimer–Volkoff equations and presents the EOS model used to relate pressure and energy density. Section \ref{sec:ANN} provides an overview of artificial neural networks, with a focus on FFNs and ResNets in the context of this work. Section \ref{sec:NN_Models} details the proposed network architectures, including dataset generation, training procedures, and evaluation methodology. Section \ref{sec:Performance} presents the results, assessing model accuracy and benchmarking performance against a traditional TOV solver. Finally, Section \ref{sec:Conclusion} summarizes the key findings and outlines directions for future research.


\section{Neutron Star Observables}\label{sec:TOV}

\subsection{Structure and Tidal Deformability}\label{subsec:TOV}

The macroscopic structure of a non-rotating, spherically symmetric neutron star is governed by the TOV equations~\cite{tolman39,oppenheimer39}, derived from the Einstein field equations under the assumption of hydrostatic equilibrium. In natural units ($G = c = 1$), these take the form:
\begin{equation}
\frac{dp}{dr} = -\frac{\left[\epsilon(r) + p(r)\right]\left[m(r) + 4\pi r^3 p(r)\right]}
{r\left[r - 2m(r)\right]}\,,
\end{equation}
\begin{equation}
\frac{dm}{dr} = 4\pi r^2 \epsilon(r)\,,
\end{equation}
where $p(r)$, $\epsilon(r)$, and $m(r)$ are the pressure, energy density, and enclosed gravitational mass at radial coordinate $r$, respectively. Given a central pressure $p_c$, these equations are integrated outward from the center with boundary conditions $m(0) = 0$ and $p(0) = p_c$, using an EOS to close the system by relating $p$ and $\epsilon$. Integration proceeds until the pressure vanishes at $r = R$, which defines the stellar radius $R$ and gravitational mass $M \equiv m(R)$.

The tidal deformability $\Lambda$ characterizes the quadrupolar deformation of a neutron star induced by an external gravitational field, and is defined as~\cite{Hinderer:2007mb, Binnington:2009bb, Damour:2009vw}:
\begin{equation}
\Lambda = \frac{2}{3}\, k_2 \left(\frac{R}{M}\right)^5\,,
\end{equation}
where $k_2$ is the second tidal Love number. This quantity is obtained by solving a first-order differential equation for $y(r)$ together with the TOV equations,
\begin{equation}
\begin{aligned}
\frac{dy}{dr} = -\frac{y(r)^2}{r} 
- \frac{y(r)}{r}\left[1 + 4\pi r^2\left(p(r) - \epsilon(r)\right)\right] \\
- \frac{r\left[5\epsilon(r) + 9p(r) + \dfrac{\epsilon(r)+p(r)}{\partial p/\partial\epsilon} 
- \dfrac{6}{4\pi r^2}\right]}{r - 2m(r)} \\
+ \frac{4\left[m(r) + 4\pi r^3 p(r)\right]^2}{r^2\left[r - 2m(r)\right]^2}\,,
\end{aligned}
\end{equation}
subject to the boundary condition $y(0) = 2$. The surface value $y_R \equiv y(R)$ then determines $k_2$ via:
\begin{equation}
\begin{aligned}
k_2 = \frac{8C^5}{5}(1-2C)^2\left[2 + 2C(y_R - 1) - y_R\right] \\
\times \Bigl\{2C\left[6 - 3y_R + 3C(5y_R - 8)\right] \\
+ 4C^3\left[13 - 11y_R + C(3y_R - 2) + 2C^2(1+y_R)\right] \\
+ 3(1-2C)^2\left[2 - y_R + 2C(y_R-1)\right]\ln(1-2C)\Bigr\}^{-1}\,,
\end{aligned}
\end{equation}
where $C \equiv M/R$ is the compactness. Together, $R$, $M$, and $k_2$ fully determine the tidal deformability $\Lambda$ for a given EOS and central pressure.

\subsection{Piecewise Polytropic EOS}\label{subsec:PPEoS}

We model the neutron star interior using a piecewise polytropic EOS (PPEOS) parameterization~\cite{Read:2008iy}, in which the EOS is divided into density segments each described by:
\begin{equation}
p(\rho) = K_i \rho^{\Gamma_i}, \qquad
\epsilon(\rho) = \frac{K_i}{\Gamma_i - 1}\rho^{\Gamma_i} + (1 + a_i)\rho\,,
\end{equation}
where the constants $K_i$ and $a_i$ are fixed by enforcing the continuity of pressure and energy density across segment boundaries. The high-density core is described by four free parameters, $\log p_1$, $\Gamma_1$, $\Gamma_2$, and $\Gamma_3$, where $p_1$ is the pressure at the first transition density, and the $\Gamma_i$ are the polytropic indices in three successive density intervals separated at $1.8\,\rho_0$ and $3.6\,\rho_0$.


\section{Neural Network Architectures}\label{sec:ANN}
%
\begin{figure}[ht!]
    \centering
    \includegraphics[
    width=\linewidth,]{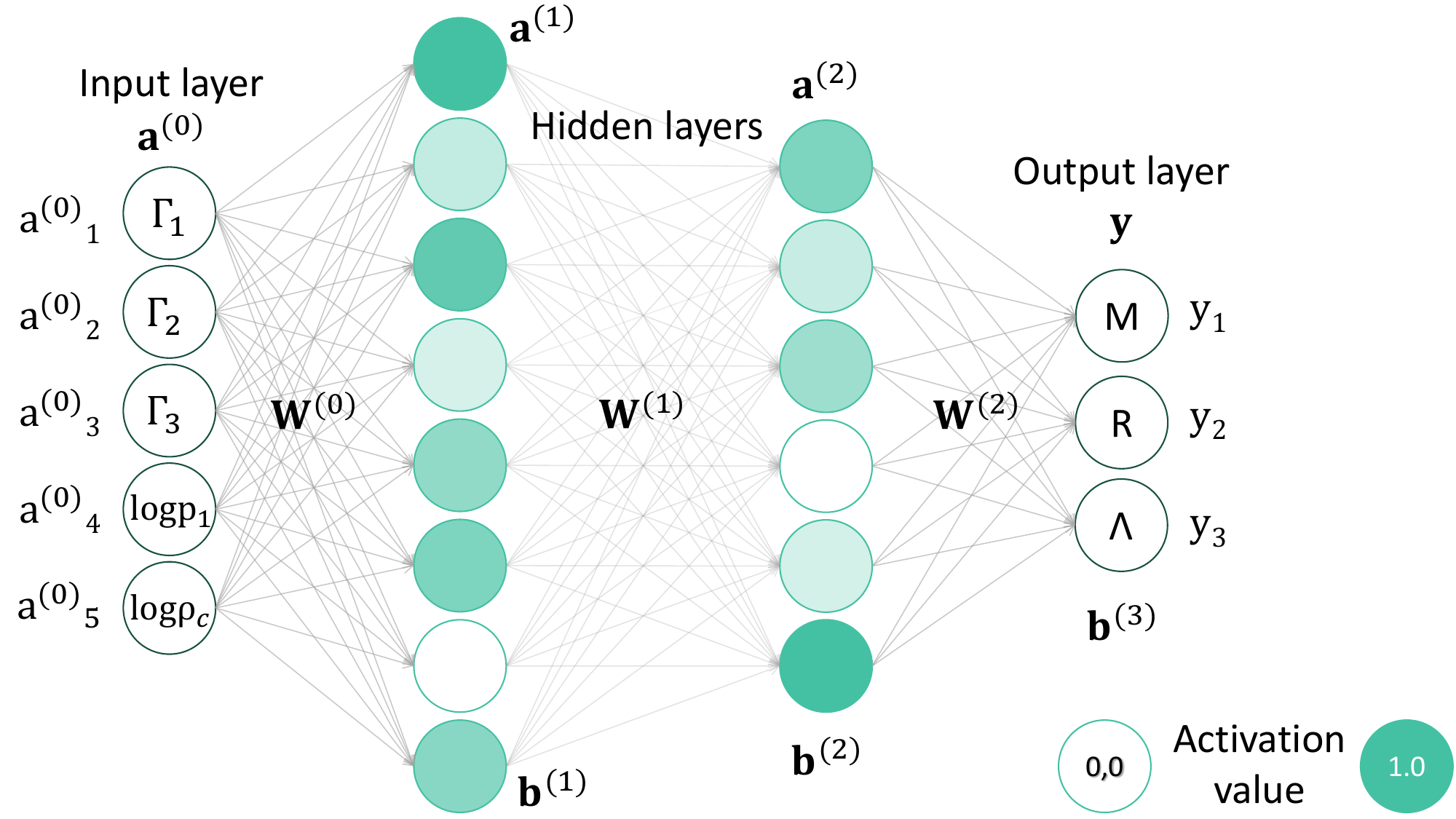}
    \caption{Schematic of a fully-connected feedforward network mapping EOS parameters $(\Gamma_1,\Gamma_2,\Gamma_3,\log p_1)$ along with  log central density $(\log \rho_c)$ inputs to neutron star observables $M$, $R$, and $\Lambda$. The output vector is given by \textbf{y}=$(M, R, \Lambda)^T$, where $M$, $R$, and $\Lambda$ denote the predicted gravitational mass, radius, and tidal deformability, respectively.
    }
    \label{fig:feed-forward_network}
\end{figure}
This section describes the architecture and working of FFNs in the context of neutron star parameter regression. We then introduce ResNets and motivate their use as an alternative architecture for the present task.

\subsection{Fully-Connected Feedforward Networks}\label{sec:FCFFN}
A fully-connected FFN maps inputs to outputs through a sequence of weighted nonlinear transformations. As illustrated in Fig.~\ref{fig:feed-forward_network}, the network consists of an input layer, two hidden layers, and an output layer. Information propagates through the network by applying an affine transformation followed by a nonlinear activation function at each hidden layer. For a generic layer $(l+1)$, the activation is given by
\begin{equation}
\mathbf{y}^{(l+1)} =
f\left(\mathbf{W}^{(l)}\mathbf{a}^{(l)}+\mathbf{b}^{(l+1)}\right),
\end{equation}
where $\mathbf{a}^{(l)}$ denotes the activations from layer $(l)$, $\mathbf{W}^{(l)}$ is the weight matrix connecting layer $(l)$ to layer $(l+1)$, $\mathbf{b}^{(l+1)}$ is the corresponding bias vector, and $f$ represents the nonlinear activation function. Common choices include the ``rectified linear unit'' (ReLU)~\cite{relu2010} and the ``Gaussian error linear unit'' (GELU)~\cite{2016arXiv160608415H}, which introduce the nonlinear representational capacity required to model complex mappings. The specific choice of activation function influences the learning behavior of the network by affecting gradient propagation, numerical stability, and the ability of the model to approximate highly nonlinear relationships.

For the input layer, $\mathbf{a}^{(0)}$ corresponds to the input features $(\Gamma_1,\Gamma_2,\Gamma_3,\log p_1,\log \rho_c)$, while the output layer produces the predicted neutron-star observables $(M,R,\Lambda)$, as shown in Fig.~\ref{fig:feed-forward_network}. During training, the cost function $C$ is evaluated after each forward pass to quantify the difference between the predicted and target outputs. The gradients of the cost function, $-\nabla C$, are then propagated backwards through the network to update the weights and biases, iteratively improving the model predictions. This process, known as backpropagation, constitutes the training procedure of the neural network~\cite{goodfellow2016deep, NN, NN_book}.

Tasks involving complex nonlinear mappings often require networks with multiple hidden layers, forming Deep Neural Networks (DNNs), which provide increased representational capacity for learning intricate relationships~\cite{DNN, Estimation, Detection}. A schematic of the fully-connected FFN architecture is shown in Fig.~\ref{fig:feed-forward_network}.

\subsection{Residual Networks}\label{subsec:ResNets}

The task considered here, namely simultaneously predicting $(M,R,\Lambda)$ from five EOS and stellar input parameters across $\mathcal{O}(10^5)$ training samples, requires a neural network with sufficient capacity to represent highly nonlinear relationships. Increasing network depth is one natural way of enhancing representational power. However, deeper architectures do not necessarily yield improved performance. Previous studies have shown that standard feed-forward networks can become increasingly difficult to optimize as depth increases, with training error sometimes saturating or degrading despite the additional model capacity \cite{ResNets,ResNet_reference_11,ResNet_reference_42} a phenomenon distinct from overfitting. Such behaviour has often been associated with optimization challenges including vanishing gradients \cite{ResNet_reference_1,ResNet_reference_9}: gradients backpropagated through successive nonlinear transformations diminish and become uninformative, causing early layers to learn ineffectively.

Although modern training techniques such as improved weight initialization \cite{ResNet_reference_9,ResNet_reference_13,ResNet_reference_23,ResNet_reference_37} and normalization layers \cite{ResNet_reference_22} can substantially alleviate these issues, residual architectures remain an attractive alternative for deep regression problems.
\begin{figure}[ht!]
    \centering
    \includegraphics[
    width=\linewidth,
    trim=6cm 4cm 4cm 5cm,
    clip]{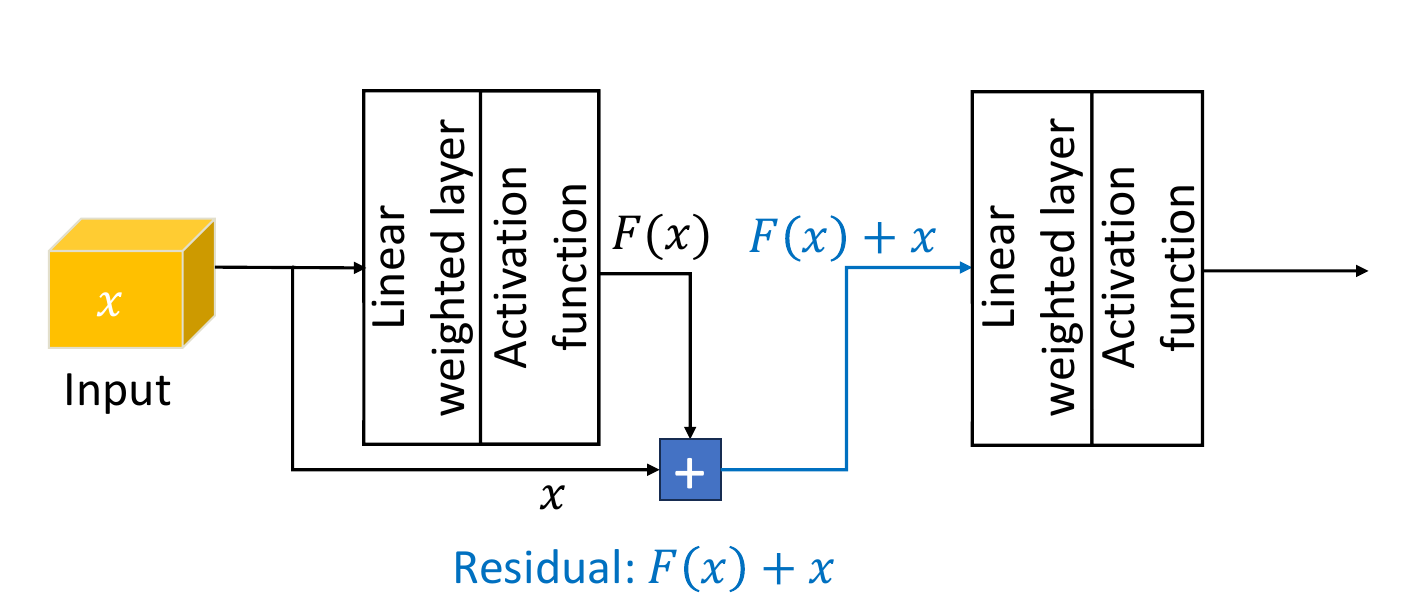}
    \caption{Residual block with identity skip connection. The input $\mathbf{x}$ 
    bypasses the nonlinear transformation $F(\mathbf{x})$ and is added directly 
    to the block output, forming the residual mapping $F(\mathbf{x}) + \mathbf{x}$.}
    \label{fig:Skip_connection}
\end{figure}
{\bf Residual Networks (ResNets)} provide an alternative mechanism for information and gradient propagation through the network by introducing skip connections ~\cite{ResNet_reference_2, ResNet_reference_34, ResNet_reference_49}, that add the block input directly to its output, as illustrated in Fig.~\ref{fig:Skip_connection}:
\begin{equation}
\mathbf{y} = F(\mathbf{x}) + \mathbf{x}\,,
\end{equation}
where $F(\mathbf{x})$ represents the transformation learned by the stacked layers and $\mathbf{x}$ is passed through unchanged via the identity shortcut. Rather than learning the full target mapping $H(\mathbf{x})$ directly, the network learns only the residual $F(\mathbf{x}) := H(\mathbf{x}) - \mathbf{x}$ that has demonstrated strong performance across a variety of machine-learning applications \cite{ResNets}.

The skip connection provides an unobstructed gradient pathway during backpropagation, ensuring that early layers receive meaningful updates regardless of network depth~\cite{ResNets, goodfellow2016deep}. Furthermore, if the optimal mapping is close to the identity, the network can represent it by driving $F(\mathbf{x}) \to 0$, substantially reducing the optimization difficulty. These properties make ResNets particularly well-suited for the high-dimensional nonlinear regression task of simultaneously predicting neutron star mass, radius, and tidal deformability from EOS parameters.

In the present work, the motivation for investigating ResNets is not that the feed-forward architecture exhibits optimization difficulties, but the motivation is rather to assess whether residual learning offers any advantage for the EOS-to-observable mapping. Given the nonlinear dependence of neutron-star observables on the underlying EOS, it is reasonable to explore whether the enhanced feature-learning capabilities of residual architectures can improve predictive accuracy, training behaviour, or computational efficiency relative to a conventional FFN.


\section{Neural Network Models}\label{sec:NN_Models}
%
\begin{figure*}[ht!]
    \centering
    \includegraphics[width=\linewidth,trim=0cm 3cm 0cm 5cm,clip]{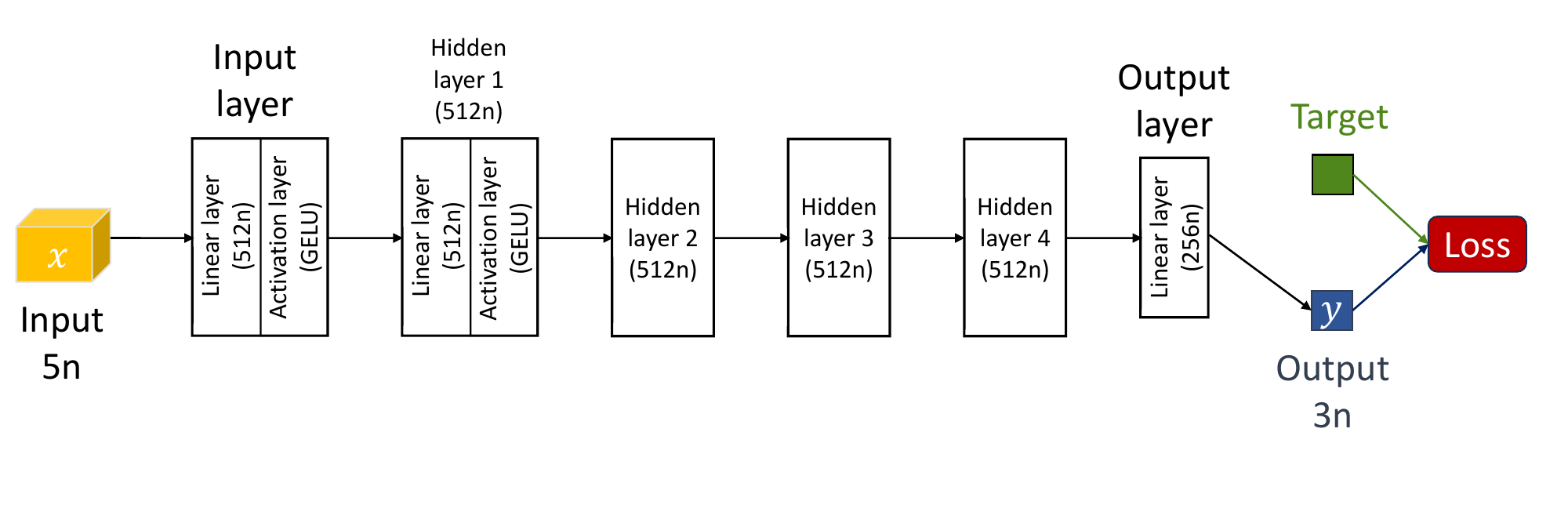}
    \caption{Architecture of the proposed feed forward neural network. Five input features are 
    encoded into a 512-dimensional representation, processed by four hidden layers, and projected onto the three neutron star observables 
    $(M, R, \log_{10}\Lambda)$.}
    \label{fig:FFN}
\end{figure*}
\begin{figure*}[ht!]
    \centering
    \includegraphics[width=\linewidth,trim=0cm 3cm 0cm 5cm,clip]{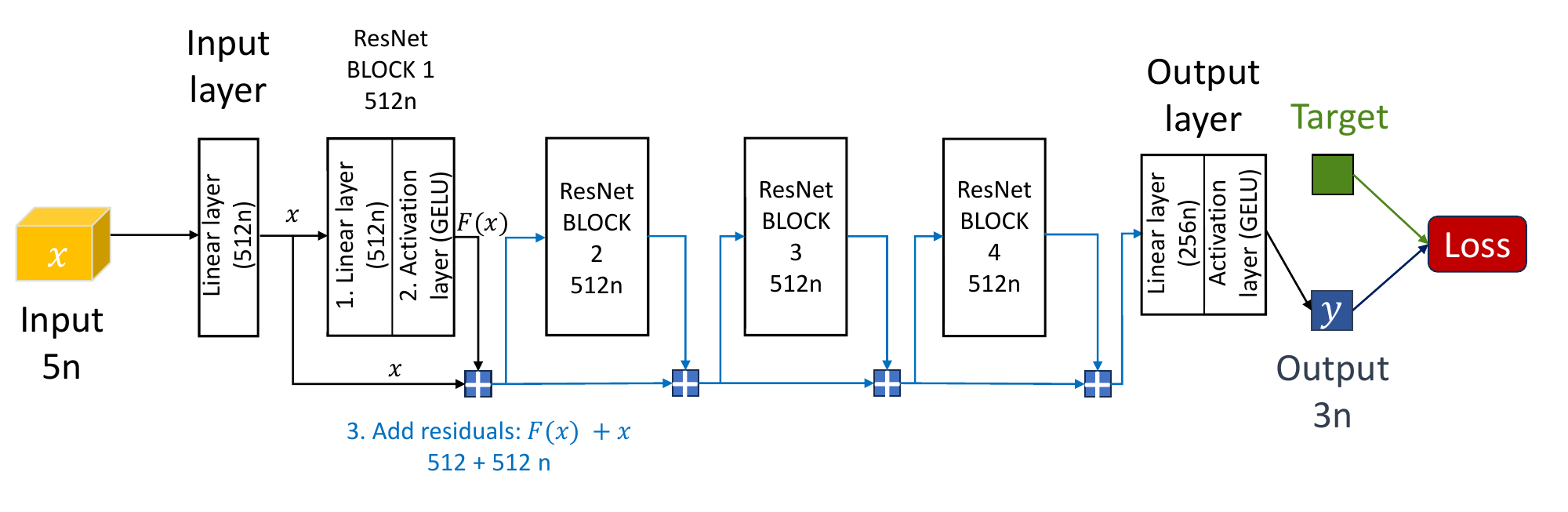}
    \caption{Architecture of the proposed residual network. Five input features are 
    encoded into a 512-dimensional representation, processed by four residual blocks with 
    identity skip connections, and projected onto the three neutron star observables 
    $(M, R, \log_{10}\Lambda)$.}
    \label{fig:ResNet}
\end{figure*}
This section describes the proposed model architectures, the construction of the training dataset, the training methodology, and the procedure used to evaluate the models on previously unseen data.

\subsection{Model Architecture}\label{subsec:Model_arch}

The proposed architectures, shown in Figs.~\ref{fig:FFN} and ~\ref{fig:ResNet}, map five input features to three neutron star observables. The inputs comprise the four PPEOS parameters defining the high-density core ($\log p_1, \Gamma_1, \Gamma_2, \Gamma_3$; see Sec.~\ref{sec:TOV}) together with the logarithmic central density $\log\rho_c$, which selects a configuration along the resulting stellar sequence. The three outputs are the gravitational mass $M$ (in $M_\odot$), the radius $R$ (in km), and the tidal deformability expressed as $\log_{10}\Lambda$ which compresses its large dynamic range and becomes comparable to the other parameters.

For the feed forward network in Fig.~\ref{fig:FFN} the five inputs are mapped onto a 512-dimensional latent representation by a linear input layer, which subsequently flows through four hidden layers with GELU activation before connecting to the output layer comprised of 256 neurons which eventually feed into the three outputs.

In the case of the ResNet, Fig.~\ref{fig:ResNet} the five inputs are first mapped to a 512-dimensional latent representation as well by a linear input layer, which feeds into a sequence of four residual blocks. Each block applies a linear transformation followed by a GELU activation, and adds the result to its input through an identity skip connection:
\begin{equation}
\mathbf{h}^{(l+1)} = \mathbf{h}^{(l)} + 
\mathrm{GELU}\!\left(\mathbf{W}^{(l)}\mathbf{h}^{(l)} + \mathbf{b}^{(l)}\right)\,,
\end{equation}
where the hidden dimension is held fixed at 512 throughout, so that the skip connection is a direct element-wise addition. The final output layer maps the 512-dimensional representation to 256 neurons through a linear layer with GELU activation, and then projects to the three output quantities; unlike the residual blocks, it does not employ a skip connection.

\subsection{Dataset Creation}\label{subsec:Dataset}

The training dataset was generated by solving the TOV and tidal perturbation equations 
(Sec.~\ref{sec:TOV}) for a large set of piecewise polytropic equations of state. The four EOS parameters and the central density were sampled uniformly over the ranges $\Gamma_{1,2,3} \in [1.4, 5.0]$, $\log p_1 \in [33.5, 34.8]$, and $\log\rho_c \in [14.5, 15.4]$, where $p_1$ is in $\mathrm{dyn\,cm^{-2}}$ and $\rho_c$ in $\mathrm{g\,cm^{-3}}$.

For each sampled configuration, the stellar structure and tidal perturbation equations were solved in Fortran using a vectorized, parallelized fourth-order Runge-Kutta (RK4) solver with a spatially adaptive step size while simultaneously computing the tidal deformability. The solver was interfaced with the Python data-generation pipeline through an F2PY~\cite{peterson2009f2py} wrapper. The integration was initialized with a step size of $h=10^{-4}~\mathrm{km}$, which is subsequently adjusted according to the local difference between the fractional radial gradients of the enclosed mass and pressure,
\begin{equation}
    \left|\frac{1}{m}\frac{dm}{dr}-\frac{1}{p}\frac{dp}{dr}\right|.
\end{equation}
This heuristic criterion decreases the step size in regions where the solution varies rapidly and increases it where the solution evolves more smoothly, providing an efficient balance between computational cost and numerical accuracy.

For each configuration, the solver returns the gravitational mass, radius, and tidal deformability. Physical consistency was enforced by requiring a monotonically increasing pressure profile and causality throughout the star, while the unstable branch beyond the maximum-mass configuration was discarded. Configurations yielding unphysical observables, for example due to extremely stiff equations of state, were removed by restricting the outputs to $M \in [0.15, 3.5]\,M_\odot$, $R \in [6, 25]\,\mathrm{km}$, and $\Lambda \in [0, 10^6]$. 
This procedure yielded $4.2 \times 10^5$ valid samples used for training and validation.

The dataset was partitioned into training and validation sets using an 80--20 split. The five input features were standardized via Z-score normalization ~\cite{stat_learning_introduction}. The outputs were rescaled to comparable dynamic ranges: mass and radius by constant factors, and the tidal deformability through the transformation $\Lambda \to \log_{10}\Lambda$. This preprocessing places all quantities on similar scales, improving numerical conditioning and convergence during training~\cite{goodfellow2016deep}.

\subsection{Training}\label{subsec:Training}

The models were trained by minimizing the Huber loss~\cite{Huber1964} between predicted and target observables, using the AdamW~\cite{2017arXiv171105101L} optimizer with a cosine-annealing learning-rate ~\cite{loshchilov2017sgdrstochasticgradientdescent}. The Huber loss was chosen for its robustness to the occasional large residual relative to the mean squared error. The full set of training hyperparameters is listed in Table~\ref{table1}.

Training was configured for a maximum of 1000 epochs, with early stopping employed to prevent overfitting~\cite{overfitting}. Convergence was reached at epoch 364, with a best validation loss of $7.45 \times 10^{-7}$, in approximately 37 minutes of wall-clock time in the case of the FFN whereas the ResNet training was stopped early at epoch 685 using the same criterion as before and reached a validation loss of $3.12 \times 10^{-7}$, in approximately 70 minutes. The training was performed on a single compute node of a HPC cluster equipped with an AMD Ryzen 9 7950X3D Processor (16 physical cores, 32 logical cores with hyper-threading enabled) and 62 GB of RAM. 
The training and validation loss curves for the FFN is shown in Fig.~\ref{fig:loss_curve_final_FFN} and the same for the ResNet is shown in Appendix~\ref{app:ResNet_training} (Fig.~\ref{fig:loss_curve_final_RN}).
\begin{figure}[ht!]
    \centering
    \includegraphics[
    width=\linewidth,
    trim= 0cm 0cm 0cm 0cm,
    clip]
    {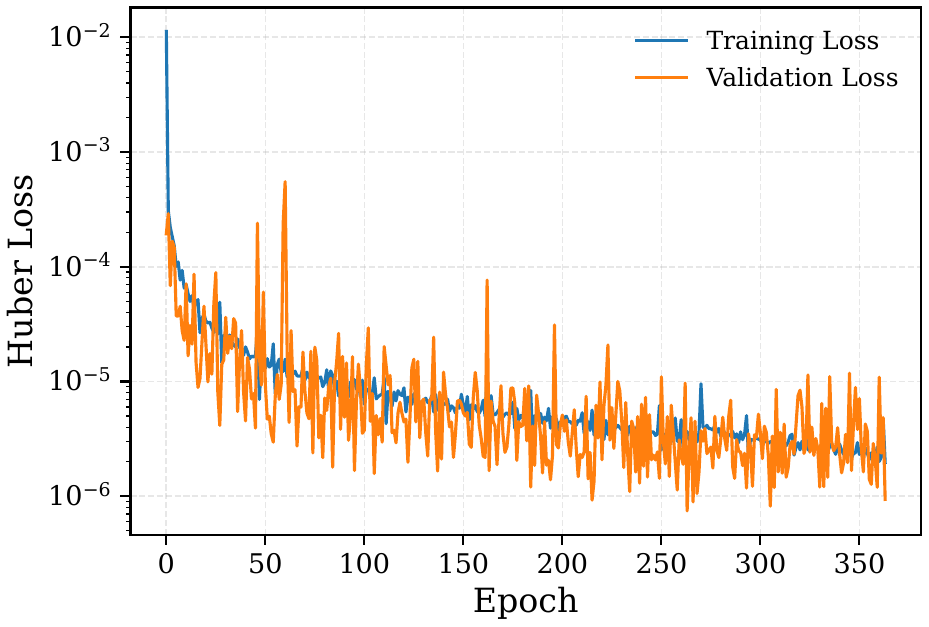}
    \caption {Training and validation loss as a function of epoch for the FFN. The close agreement between the two curves indicates the absence of significant overfitting.}
    \label{fig:loss_curve_final_FFN}
\end{figure}
\begin{table}[ht!]
\caption{Training hyperparameters of the models.}
\label{table1}
\begin{tabular}{lr}
\hline\hline
    Batch size & 256 \\
    Learning rate & $5\times10^{-4}$ \\
    Optimizer & AdamW \\
    Loss function & Huber loss \\
    Activation function & GELU \\
    Maximum epochs & 1000 \\
    Early-stopping patience & 100 epochs \\
\hline\hline
\end{tabular}
\end{table}
%
\subsection{Testing}\label{subsec:Testing}

The trained models were evaluated on EOSs not seen during training. Test configurations were drawn by selecting combinations of $(\Gamma_1, \Gamma_2, \Gamma_3, \log p_1)$ within the training ranges, since predictions outside these ranges would require extrapolation beyond the regime the models were trained on. For each set of EOS parameters, a sequence of stellar models was generated by varying the central density $\log\rho_c$.

For every configuration, the mass, radius, and tidal deformability were computed directly from the TOV and tidal equations to provide ground-truth reference values, and were also predicted by the trained networks. Reference configurations were retained only where the numerical solution returned a physically valid, stable star; predictions were excluded only when numerically non-finite, and were not filtered on their physical values, so that the reported accuracy reflects the model's performance without bias. 
Agreement between predicted and reference values was assessed through the mass--radius and tidal deformability--mass relations, and quantified using the mean absolute error (MAE), root-mean-square error (RMSE), combined normalized RMSE, and the coefficient of determination ($R^2$). In addition, the wall-clock times of the network and the numerical solver were compared to quantify the computational speedup, as detailed in Sec.~\ref{sec:Performance}.


\section{Model Performance}\label{sec:Performance}

\subsection{Accuracy}\label{subsec:Accuracy}

The accuracy of the trained models was assessed on EOSs not seen during training, using the MAE, RMSE, combined normalized RMSE, and $R^2$ metrics introduced in Sec.~\ref{subsec:Testing}. These metrics are summarized in Table~\ref{table2}, for a test set of 100 distinct EOSs, with each EOS sampled at 100 central densities. Additional tests performed on 1000 distinct EOSs produced nearly identical results and are therefore omitted in the table, serving instead as a reproducibility check of the reported performance.

The combined normalized RMSE across all three observables for 100 EOSs, with each EOS being sampled at 100 central densities is $2.106\times10^{-3}$ in the case of the FFN and $0.449\times10^{-3}$ in the case of the ResNet.
In all cases the coefficient of determination exceeds $0.999$, indicating that the network reproduces the numerical TOV solutions to high accuracy across the full parameter range. The close agreement between the results obtained using 100 and 1000 EOSs further demonstrates the robustness and reproducibility of the model's predictive performance.
\begin{figure}[!t]
    \centering
    \includegraphics[width=\linewidth, trim= 0cm 0cm 0cm 0cm, clip]
    {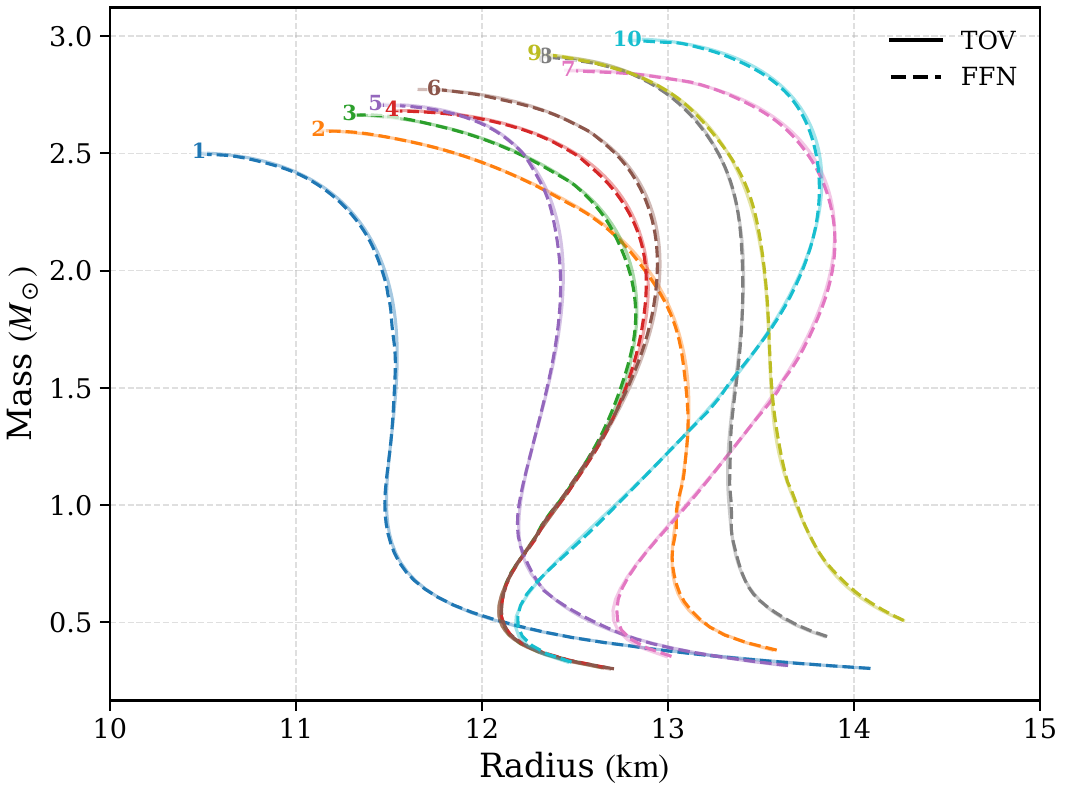}
    \caption{Mass--radius relation for 10 unseen equations of state, each evaluated at 
    100 central densities. The numerical TOV solution (solid) and the FFN prediction 
    (dashed) are shown for comparison.}   
    \label{fig:MR_unseen_FFN}
\end{figure}
\begin{figure}[!t]
    \centering
    \includegraphics[width=\linewidth, trim= 0cm 0cm 0cm 0cm, clip]
    {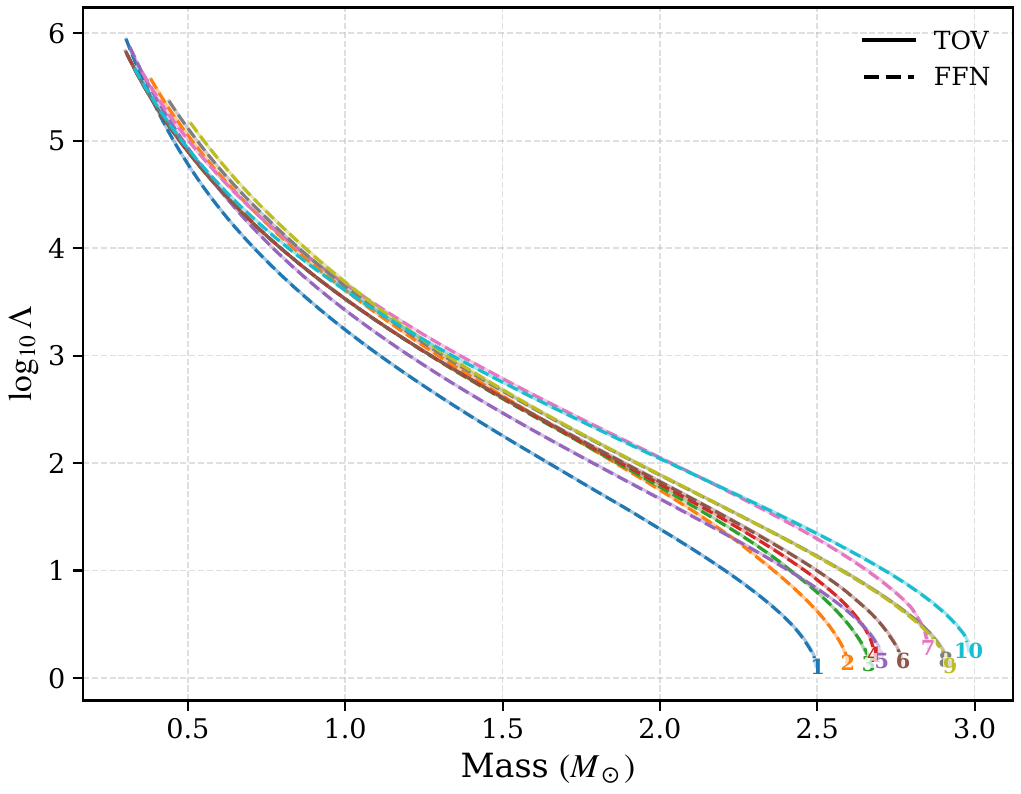}
    \caption{Tidal deformability as a function of mass for the same 10 unseen equations 
    of state. The numerical TOV solution (solid) and the FFN prediction (dashed) are 
    shown for comparison.}
    \label{fig:LM_unseen_FFN}
\end{figure}

Figure~\ref{fig:MR_unseen_FFN} compares the mass--radius relation computed by the numerical solver (solid) with the FFN prediction (dashed) for 10 unseen equations of state, each evaluated at 100 central densities. The predicted sequences track the reference curves closely across the entire mass range, including the vicinity of the maximum mass. The corresponding PPEOS parameter values for the selected test cases are provided in Appendix~\ref{app:EOS_Table}. Figure~\ref{fig:LM_unseen_FFN} shows the corresponding tidal-deformability--mass relation for the same equations of state, where the predictions again closely follow the reference values. Figures ~\ref{fig:MR_unseen_RN} and ~\ref{fig:LM_unseen_RN} shown in Appendix ~\ref{app:ResNet_testing} represent the same predictive power in the case of the ResNet model, while the EOS parameters used for these cases are also listed in Appendix~\ref{app:EOS_Table}.
\begin{figure}[!t]
    \centering
    \includegraphics[width=\linewidth, trim= 0cm 0cm 0cm 0cm, clip]
    {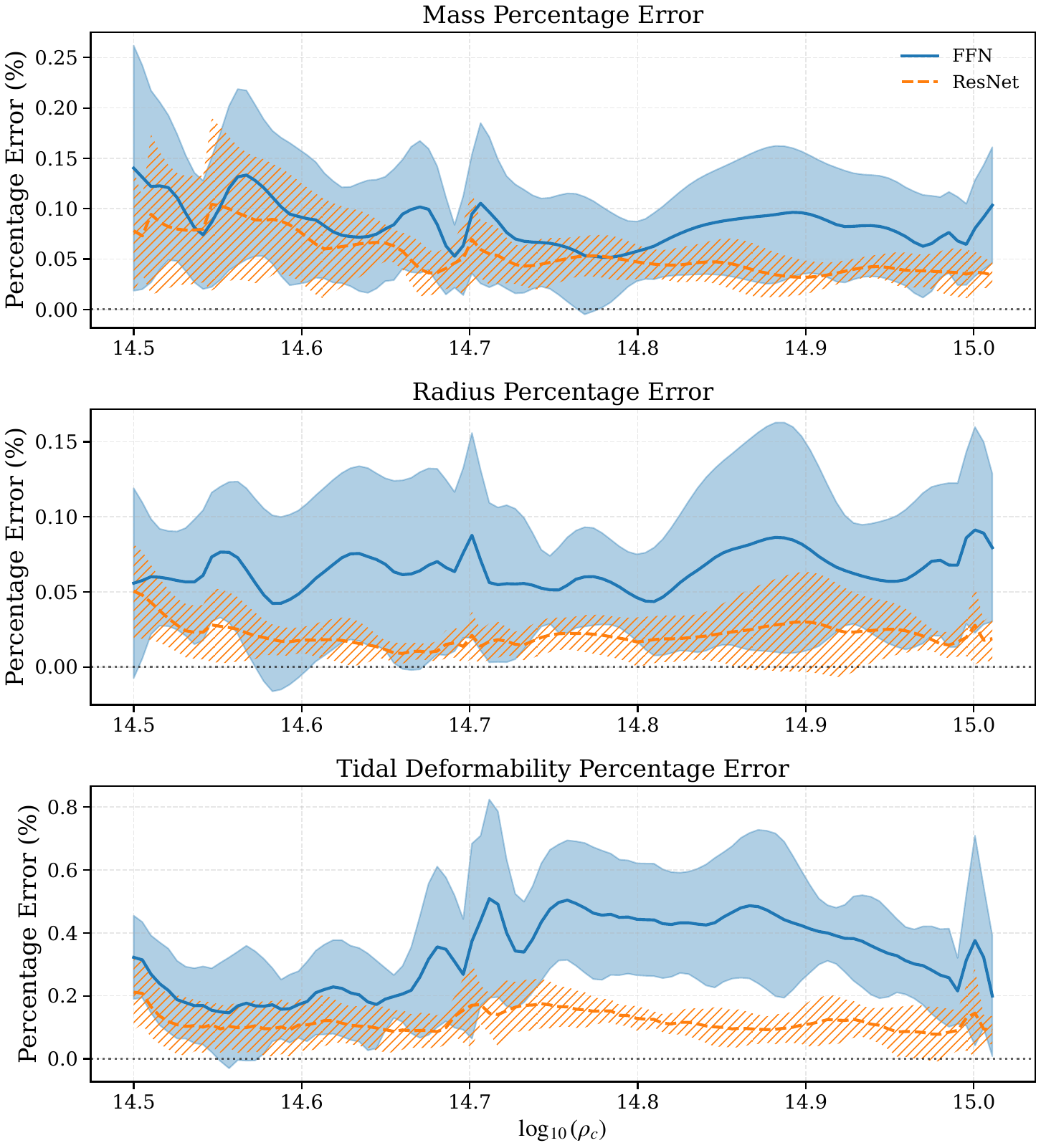}
    \caption{Aggregate percentage errors for Mass, Radius and Tidal Deformability vs logarithmic central density for 10 unseen equations of state, each evaluated at 100 central densities.
    }
    \label{fig:agg_error}
\end{figure}

To quantify the predictive performance across the full stable branch, the relative prediction error was evaluated as a function of logarithmic central density. Figure ~\ref{fig:agg_error} presents the aggregate percentage error between predicted and reference values: the mass, radius and  tidal-deformability values against the logarithmic central density. Since the maximum stable central density differs between EOSs, all stellar sequences were first interpolated onto a common logarithmic central-density grid spanning from $\log_{10}(\rho_c)=14.5$ to the smallest maximum stable density among the sampled EOSs. This common grid ensures that errors from different EOSs are compared at identical central densities while avoiding extrapolation beyond the stable branch.

For each EOS, the pointwise relative error was computed as: 
\begin{equation}
    \begin{aligned}
        \epsilon(\rho_c)=\frac{|X_{pred} - X_{true}|}{X_{true}}
    \end{aligned},
\end{equation}
where X denotes the stellar mass, radius or tidal deformability. 
The interpolated errors from all EOSs were then combined to calculate the mean prediction error and the corresponding one-standard-deviation uncertainty at each central density. This provides an aggregate measure of the emulator's performance across the entire ensemble of sampled equations of state. 

The resulting profiles reveal how the emulator accuracy varies across the stable branch, from lower-mass stars at smaller central densities to configurations approaching the maximum stable mass. Regions where the mean error increases indicate areas of greater predictive difficulty, while the width of the one-standard-deviation band reflects the sensitivity of the predictions to variations in the underlying equation of state. The percentage errors remain small and unbiased across the sampled central-density range for both models, with no significant difference in their predictive performance. It should also be noted that the vertical axis spans a narrow range of percentage errors, which visually exaggerates the apparent width of the uncertainty bands. Overall, the figure demonstrates that both NN architectures maintain consistently high predictive accuracy across the full range of physically stable NS configurations.
\begin{table}[tbp]
\caption{Model metrics for 100 EOSs each with 100 stars when predicting mass, radius, and tidal deformability. MAE and RMSE values are in units of $10^{-3}$.}
\label{table2}
\centering
\makegapedcells
\renewcommand{\tabcolsep}{5pt}
\begin{tabular}{c c c c c}
\hline\hline
\thead{Model \\ type} & \thead{Metric} & Mass & Radius & \thead{Tidal \\ Def.} \\
\hline
\multirow[c]{3.4}{*}{FFN} 
  & MAE  & $2.351$ & $11.827$ & $1.234$ \\
  & RMSE & $5.720$ & $23.901$ & $1.984$ \\
  & $R^2$ & $0.999936$ & $0.999622$ & $0.999998$ \\\\
  
\multirow[c]{3.4}{*}{ResNet} 
  & MAE  & $0.769$ & $3.151$ & $0.366$ \\
  & RMSE & $1.012$ & $5.199$ & $0.555$ \\
  & $R^2$ & $0.999998$ & $0.999983$ & $0.999999$ \\
\hline\hline
\end{tabular}
\end{table}
%

\subsection{Benchmarks}\label{subsec:Benchmarks}

The computational performance of the networks was benchmarked against the numerical TOV solver used to generate the training data. The solver performs direct fourth-order Runge--Kutta integration of the TOV and tidal equations for each requested stellar configuration. Benchmarks were run for 100, 1000, and 10,000 equations of state, each evaluated over 100 central densities. All timing measurements were obtained on the same compute node used for training, with the numerical solver executed on the CPU and the neural networks performing inference on the same hardware.

To ensure a like-for-like comparison, the networks were evaluated only on the stable stellar configurations identified by the TOV solver. Consequently, the TOV runtime includes the computational cost of identifying and validating the stable branch, whereas the neural network performs inference directly on this validated set of stellar models.

To reduce initialization bias, the first execution of both methods was treated as a warm-up and discarded. Each configuration was then repeated ten times and the median runtime recorded, the median being more robust than the mean against transient fluctuations from operating-system scheduling and background activity. The standard deviation across runs was retained as a measure of timing stability. The speedup factor is defined as
\begin{equation}
S = \frac{t_{\mathrm{TOV}}}{t_{\mathrm{model}}}\,,
\end{equation}
the ratio of solver to network runtimes.

The results, listed in Table~\ref{table3}, show that the inference time of both surrogate models remains approximately constant at $\sim3.5$~ms across all benchmark sets (each EOS sampled at 100 central densities). In contrast, direct numerical integration of the TOV equations requires $\sim0.8$~s per evaluation. Both architectures therefore achieve speedups exceeding $200$, with the FFN reaching maximum median accelerations  by factors of $238.6$ and 220 for the ResNet; the slightly lower ResNet speedup reflects the additional computation associated with the residual connections. The run-to-run standard deviations remain below about $1\%$ of the measured speedup, and decrease with increasing EOS count, confirming that the timings are reproducible and largely free of background interference. Overall, the networks deliver a substantial and stable acceleration over direct numerical integration, making them well-suited to large-scale Bayesian inference and population studies in which repeated TOV evaluations would otherwise dominate the computational cost.
\begin{table}[tbp]
\caption{Runtime comparison between the numerical TOV solver and the networks for predicting mass, radius, and tidal deformability.}
\label{table3}
\centering
\makegapedcells
\renewcommand{\tabcolsep}{5pt}
\begin{tabular}{c c c c c}
\hline\hline
\thead {Network \\ type} & \thead{No. of \\ EoS} & \thead{Median \\ TOV time \\ (ms)} & \thead{Median \\ model time \\ (ms)} & \thead{Speedup \\ factor} \\
\hline
\multirow{3.4}{*}{FFN} 
    & 100 & 824.358  & 3.506  & $235.3\times \pm 2.5\times$\\
    & 1000 & 830.470  & 3.527  & $235.5\times \pm 1.3\times$\\
    & 10000 & 830.645  & 3.481  & $238.6\times \pm 0.4\times$\\\\

\multirow{3.4}{*}{ResNet} 
    & 100 & 760.101  & 3.627  & $209.6\times \pm 2.5\times$\\
    & 1000 & 762.516  & 3.503  & $217.7\times \pm 0.7\times$\\
    & 10000 & 764.513  & 3.475  & $220.0\times \pm 0.4\times$\\
\hline\hline
\end{tabular}
\end{table}
%

\section{Conclusion}\label{sec:Conclusion}

In this work, we developed and evaluated feed-forward and deep residual neural networks as surrogate models for the forward TOV mapping, predicting neutron-star mass, radius, and tidal deformability directly from the equation-of-state parameters and central density. To our knowledge, this is the first study to investigate residual learning for TOV surrogate modeling. Both architectures reproduce the numerical TOV solutions with high accuracy across the full parameter range, achieving $R^2$ values exceeding $0.999$ for all three observables.

The surrogate models accelerate the evaluation of neutron-star observables by approximately two orders of magnitude relative to direct numerical integration, with measured speedups ranging from approximately $209\times$ to $238\times$, depending on the network architecture and sampling density.

We note that a direct comparison with previously reported emulators is complicated by differences in the numerical solver, hardware, and the set of predicted observables. For instance, Deep TOV~\cite{Tiwari:2024jui} reported an approximate $8\times$ acceleration over the Python interface to the TOV solver provided by the RePrimAnd package~\cite{Kastaun:2020uxr} for predicting mass and radius on a single CPU core.
In the present work, the benchmark is performed against a vectorized, parallelized RK4 solver and additionally predicts the tidal deformability. Consequently, the reported speedups are not directly comparable, but they demonstrate that neural-network surrogates can provide substantial acceleration even when compared against an optimized numerical baseline.

Residual Networks were investigated as an alternative to conventional Feed-Forward Networks because of their demonstrated success in training deeper architectures and modeling highly nonlinear relationships. Given the complex dependence of neutron-star observables on the underlying equation of state through the TOV equations, residual learning was expected to potentially improve predictive performance or training behaviour. The ResNet architecture achieved marginally higher predictive accuracy than the FFN across all predicted observables, demonstrating that residual learning can provide a modest improvement in regression performance. However, this increase in accuracy came at the cost of slightly longer inference times, resulting in a somewhat smaller computational speedup relative to the TOV solver. 

The performance differences between the two architectures nevertheless remained small, indicating that the FFN already possesses sufficient representational capacity for the present EOS-to-observable mapping and that the additional complexity introduced by residual connections yields only incremental gains. This comparative study demonstrates that the surrogate's predictive capability is robust with respect to architectural choice while establishing a performance baseline for future investigations involving more complex EOS parameterizations or higher-dimensional regression tasks, where the advantages of residual learning may become more pronounced.

This level of acceleration is particularly relevant for multimessenger applications such as gravitational-wave parameter estimation and Bayesian EOS inference, where the TOV equations must be solved for large numbers of samples. The framework generalizes naturally across the piecewise-polytropic parameter space and provides a flexible foundation for future extensions to alternative EOS parameterizations and more sophisticated neutron-star models.


\appendix
\section{ResNet Training Curves}\label{app:ResNet_training}
%
\begin{figure}[ht!]
    \centering
    \includegraphics[width=\linewidth]{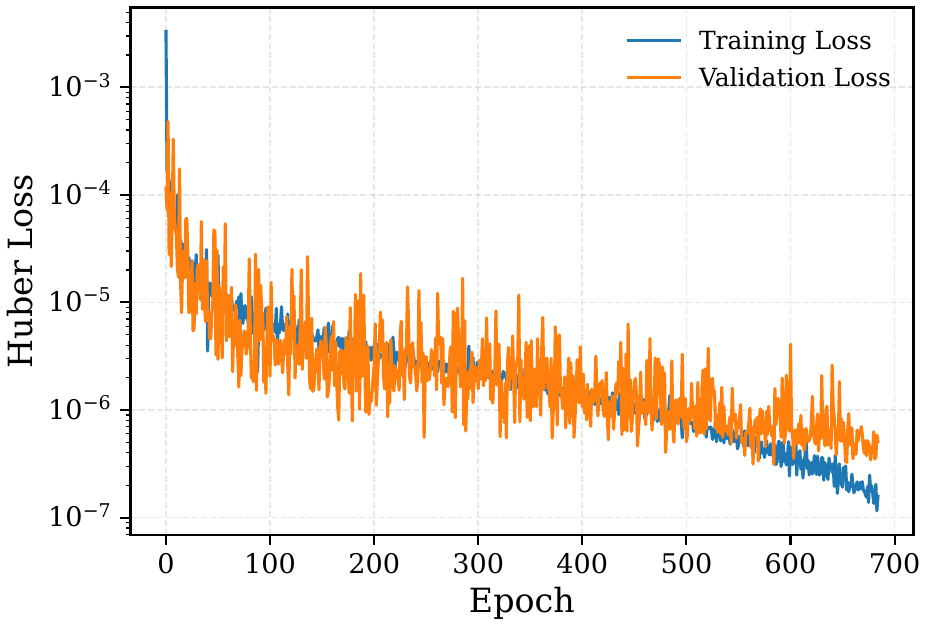}
    \caption{Training and validation loss as a function of epoch for the ResNet model.}
    \label{fig:loss_curve_final_RN}
\end{figure}
This appendix presents the training history of the ResNet model. Figure~\ref{fig:loss_curve_final_RN} shows the evolution of the training and validation Huber loss throughout the optimization process. The training and validation loss curves remain closely aligned throughout training. Toward the end of the optimization, the validation loss increases slightly above the training loss, indicating the onset of mild overfitting. However, the learning rate scheduler and early stopping terminate training before this divergence becomes significant, resulting in a model that maintains good generalization to unseen data.

\section{ResNet Testing}\label{app:ResNet_testing}
%
\begin{figure}[ht!]
    \centering
    \includegraphics[width=\linewidth]{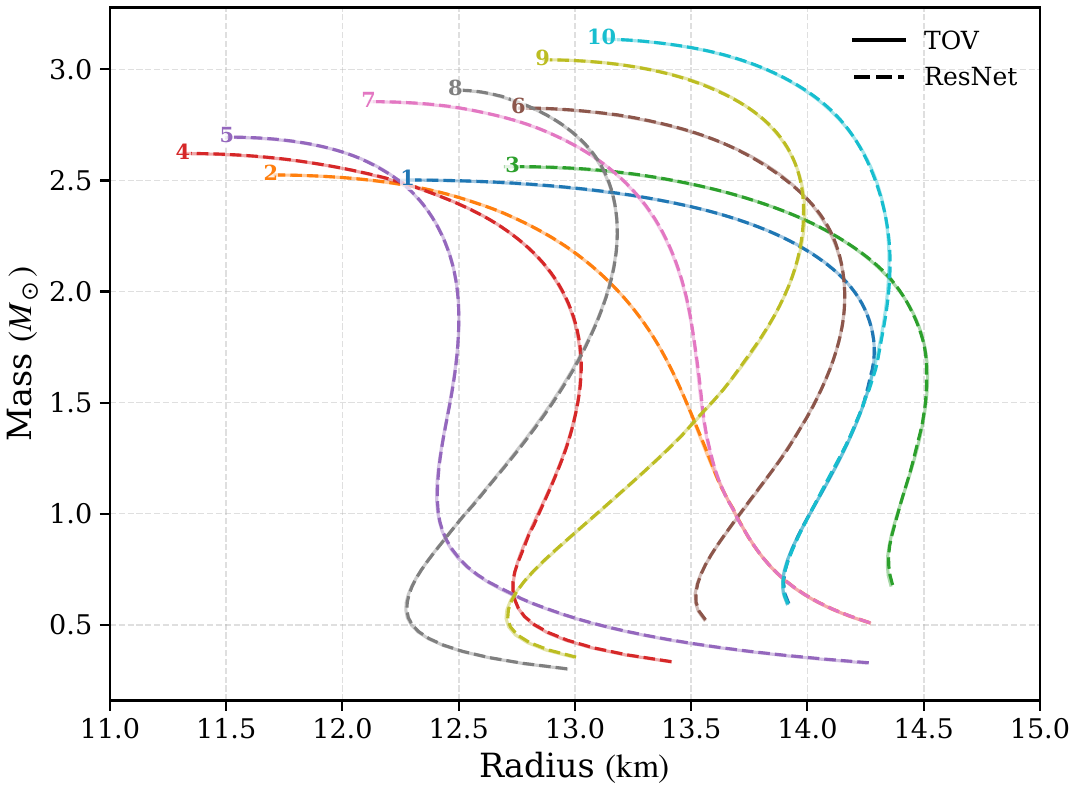}
    \caption{Mass--radius relation for 10 unseen equations of state, each evaluated at 
    100 central densities. The numerical TOV solution (solid) and the ResNet prediction 
    (dashed) are shown for comparison.}
    \label{fig:MR_unseen_RN}
\end{figure}
\begin{figure}[ht!]
    \centering
    \includegraphics[width=\linewidth]{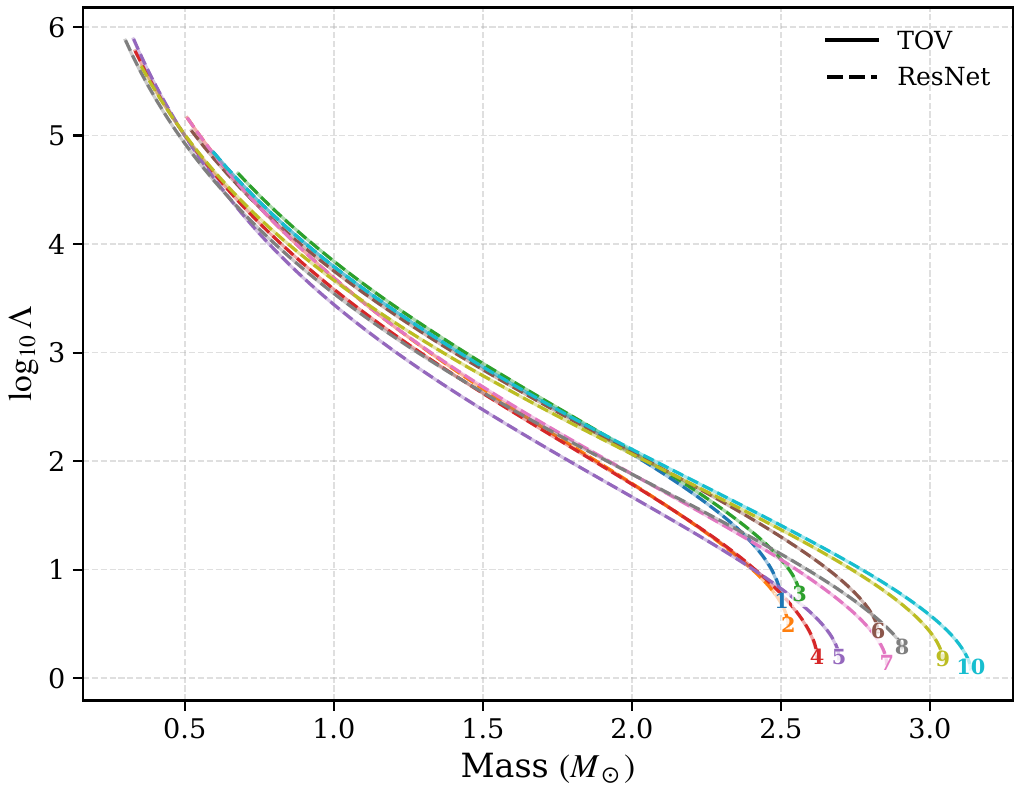}
    \caption{Tidal deformability as a function of mass for the same 10 unseen equations 
    of state. The numerical TOV solution (solid) and the ResNet prediction (dashed) are 
    shown for comparison.}
    \label{fig:LM_unseen_RN}
\end{figure}

This appendix presents the corresponding testing results for the ResNet model. The mass-radius and tidal deformability-mass diagrams, shows in figures ~\ref{fig:MR_unseen_RN} and ~\ref{fig:LM_unseen_RN}, compare the network predictions with the solutions generated by the TOV solver for the test EOS. Together with the quantitative evaluation discussed in the main text, these figures provide a visual assessment of the ResNet model's ability to reproduce the underlying physical relationships between NN observables across the full range of the test dataset.
%

\section{Equation of State Parameters for Representative Test Cases}\label{app:EOS_Table}

This appendix lists the PPEOS parameters for the representative test cases used in the mass-radius ($M-R$) and tidal deformability-mass ($\Lambda-M$) comparisons presented in Sec.~\ref{subsec:Accuracy} and Appendix ~\ref{app:ResNet_testing}. Table~\ref{tab:eos_ffn_params} gives the EOS parameter values for the curves shown for the FFN, while Table~\ref{tab:eos_resnet_params} provides the corresponding values for the ResNet. The tabulated parameters, $\Gamma_1, \Gamma_2, \Gamma_3,$ and $\log p_1$, fully specify the piecewise polytropic EOS used to generate the reference TOV solutions and are included to facilitate reproducibility.

\begin{table}[ht!]
\caption{PPEOS parameters for the representative test cases shown in the mass--radius and tidal deformability--mass curves for the FFN.}
\label{tab:eos_ffn_params}
\centering
\makegapedcells
\renewcommand{\tabcolsep}{5pt}
\begin{tabular}{ccccc}
\hline\hline
\textbf{EOS} & $\bm {\log p_1}$ & $\boldsymbol{\Gamma_1}$ & $\boldsymbol{\Gamma_2}$ & $\boldsymbol{\Gamma_3}$ \\
\hline
1 & 34.2 & 2.2 & 4.2 & 4.4 \\
2 & 34.6 & 3.2 & 3.2 & 4.8 \\
3 & 34.6 & 4.0 & 3.4 & 4.8 \\
4 & 34.6 & 4.0 & 3.6 & 4.0 \\
5 & 34.4 & 2.8 & 4.4 & 3.8 \\
6 & 34.6 & 4.0 & 3.8 & 4.6 \\
7 & 34.8 & 4.2 & 3.4 & 4.2 \\
8 & 34.6 & 3.0 & 4.4 & 3.2 \\
9 & 34.6 & 2.8 & 4.4 & 4.8 \\
10 & 34.8 & 4.8 & 3.8 & 3.0 \\
\hline\hline
\end{tabular}
\end{table}

\begin{table}[ht!]
\caption{PPEOS parameters for the representative test cases shown in the mass--radius and tidal deformability--mass curves for the ResNet.}
\label{tab:eos_resnet_params}
\centering
\makegapedcells
\renewcommand{\tabcolsep}{5pt}
\begin{tabular}{ccccc}
\hline\hline
\textbf{EOS} & $\bm{\log p_1}$ & $\boldsymbol{\Gamma_1}$ & $\boldsymbol{\Gamma_2}$ & $\boldsymbol{\Gamma_3}$ \\
\hline
1 & 34.8 & 3.4 & 2.6 & 3.0 \\
2 & 34.6 & 2.8 & 3.4 & 3.0 \\
3 & 34.8 & 3.2 & 2.8 & 2.4 \\
4 & 34.6 & 3.4 & 3.4 & 4.2 \\
5 & 34.4 & 2.6 & 4.4 & 3.6 \\
6 & 34.8 & 3.6 & 3.4 & 2.8 \\
7 & 34.6 & 2.8 & 4.2 & 4.0 \\
8 & 34.6 & 3.8 & 4.4 & 1.6 \\
9 & 34.8 & 4.2 & 4.0 & 3.4 \\
10 & 34.8 & 3.4 & 4.4 & 4.6 \\
\hline\hline
\end{tabular}
\end{table}

\section*{Acknowledgements}

BB and SR acknowledge the support from the Knut and Alice Wallenberg Foundation under grant Dnr.~KAW~2019.0112, the Deutsche Forschungsgemeinschaft (DFG, German Research Foundation) under Germany's Excellence Strategy – EXC~2121 ``Quantum Universe'' – 390833306. BB was further supported by the Alexander von Humboldt Foundation through a Humboldt Research Fellowship for Postdoctoral Researchers. SR has additionally been supported by the Swedish Research Council (VR) under grant number 2020-05044  and by the European Research Council (ERC) Advanced Grant INSPIRATION under the European Union’s Horizon 2020 research and innovation programme (Grant agreement No. 101053985).


\bibliographystyle{apsrev4-2}
\bibliography{Bibliography,mybiblio}

\end{document}